\documentclass[]{article}
\newcommand{\bb}{\begin{eqnarray}}
\newcommand{\ee}{\end{eqnarray}}
\begin{document}
\title{Quantum and classical areas of black hole thermodynamics}
\author{A. Ghosh\footnote{amit.ghosh@saha.ac.in} {}
and P. Mitra\footnote{parthasarathi.mitra@saha.ac.in}\\
Saha Institute of Nuclear Physics\\ 1/AF Bidhannagar\\
Calcutta 700064}

\date{}
\maketitle
\abstract
{
Most calculations of black hole entropy in loop quantum gravity indicate a
term proportional to the area eigenvalue $A$ with a correction
involving the logarithm of $A$.
This violates the additivity of the entropy.
An entropy proportional to
$A$, with a correction term involving the logarithm
of the classical area $k$, which is
consistent with the additivity of entropy, is derived in both U(1) and SU(2)
formulations.
}
\flushbottom
\section{Introduction}
A black hole horizon hides what is inside. This lack of information was
interpreted early on as a sign of an entropy \cite{bek}.
The interpretation was strengthened by the fact that
the area of a black hole horizon tends to increase, just like an entropy. 
These ideas led to black hole thermodynamics. 
Later, the calculation of the temperature at which black holes radiate
in quantum theory \cite{hawk} determined the scale of the entropy $S$
in terms of the area $A$ and led to a precise expression 
\bb S=\frac{A}{4\hbar G}\ee
by integrating the first law of thermodynamics.

While it may have appeared surprising that the entropy is a function of the
area instead of the volume, as happens in the case of gases, it must be
remembered that areas, like volumes, are also additive.
Since state spaces are multiplicative,
one may quite generally argue that if the number of states of a black hole
is to depend only on the area of the horizon,
\bb
N(A_1)N(A_2)=N(A_1+A_2).\label{1}
\ee
This can hold if one considers a composite black hole obtained from
two widely separated black holes, so that the horizon  has two pieces.
The equation implies that
\bb
N(A)=e^{\lambda A/2},
\ee
with some constant $\lambda$, implying the area law for the entropy $\log N$.
Nothing can be said about $\lambda$ without a microscopic approach.
It should be clear that the temperature of black hole radiation is not
an input in this argument. 
Simply the additive nature of the entropy and the area 
is used. 

Microscopic theories of black holes are supposed to tell us how this entropy
arises through a counting of states. This has been achieved to a reasonable
degree by loop quantum gravity \cite{ash,thiemann}. 
Here, the horizon area is described by an operator, with different eigenvalues.
The entropy has been found to have a term linear in this area,
with an arbitrary proportionality constant, as well as a logarithmic term,
which violates the area law \cite{bhe1,bhe2,cor}. 
We shall reanalyze the problem carefully and see how
the correct answer becomes consistent with the additivity of entropy.

The horizon in the microscopic theory of gravity is supposed to have
a set of {\it punctures} on it. Each puncture is labelled 
by spin quantum numbers $j,m$.
The area operator has eigenvalues
\bb 2\sum_p\sqrt{j_p(j_p+1)}=A,\ee 
where $j_p$ is the angular momentum quantum number associated with
the puncture $p$. This is in a special unit where
$$ 4\pi\gamma\ell_P^2=1,$$
$\gamma$ being what is called the Immirzi parameter and $\ell_P=\sqrt{\hbar G}$.
The total area of the horizon is obtained by adding the areas
contributed by the punctures. 
There is also a holonomy operator whose eigenvalues are of the form
$\exp(2\pi i a_p/k)$, where $a_p=1,2,...k$ and the integer $k$
is a measure of the classical area of the horizon (in the unit
introduced above) which is also the level of a Chern-Simons theory
describing the quantum theory of the horizon \cite{ash}. This has
to be distinguished from the quantum area $A$ arising from the punctures.
The integer $a_p$, defined modulo $k$, is required to equal
-2 times the angular momentum projection $m_p$ according to the theory.
The total spin projection then
has to satisfy
\bb \sum_p m_p=0 {\rm~mod~} k/2 \ee
so that the product of the holonomies is unity. 
We shall investigate the entropy in this theory 
using Meissner's approach \cite{bhe1} but paying attention to $k$
and find a different, $k$-dependent
logarithmic correction consistent with the additivity of entropy.
We shall also generalize that approach to the SU(2) formulation
of loop quantum gravity.


\section{U(1) loop quantum gravity \`{a} la Meissner}

Meissner's first recursion relation, where the projection constraint is
ignored, may be written as 
\bb N(A)=\sum_j
\sum_mN\Big(A-2\sqrt{j(j+1)}\Big) + \sum_m\theta(j_{max}-|m|).\ee
To understand this, one has to split the set of all distributions
of spins into those with only one puncture and those with at least two 
punctures.
If there is only one puncture, its $j$ must satisfy
$2\sqrt{j_{max}(j_{max}+1)}=A$, from which $j_{max}$ can be found.
When there are at least two
punctures, the configurations are again split, depending on the spin $j$
on the first puncture. 
This puncture can be filled with different values of $m$ for the given $j$,
{\it i.e.,} in $\sum_m 1$ ways, while
the remainder can be filled in $N\Big(A-2\sqrt{j(j+1)}\Big)$ ways
because one puncture has taken away a piece $2\sqrt{j(j+1)}$ from $A$.
The spin $j$ of the first puncture must then be summed over.
It may be mentioned that one would expect that $\sum_m 1=2j+1$,
but the value 2 has been used \cite{bhe1} in this context.
Our analysis allows both situations.

It is important to understand that this relation differs from (\ref{1})
because the area is an operator in loop quantum gravity. The simpler
relation can only hold when a unique area can be assigned to a
physical system like a black hole.

The current recursion relation is
satisfied for large $A$ by an exponential form
\bb N(A)\sim\exp(\lambda A/2)\ee with
$\lambda$ satisfying the equation 
\bb\sum_j\sum_m\exp\Big
[-\lambda\sqrt{j(j+1)}\Big]=1.\label{lambda}\ee
This fixes $\lambda$, which in turn fixes the parameter $\gamma$ required
for reproducing $S=\frac{A}{4\ell_P^2}$.

But one must impose the constraint on angular momentum projection:
the number of configurations will be reduced, so that a
correction is expected to emerge. 
One has to introduce the projection $M$ and 
the reduced number $N(A,M)$ of states.
Here $M$ stands for $\sum m$, and $M=0$ is nominally the case of interest.

Let us take a Fourier transform:
\bb N(A,M)=\int_{-2\pi}^{2\pi}e^{-i\omega M}{d\omega\over 4\pi}\tilde N(A,\omega).\ee 
For vanishing spin projection,
\bb N(A,0)=\int_{-2\pi}^{2\pi}{d\omega\over 4\pi}\tilde N(A,\omega).\ee
The recursion relation involving $M$ is 
\bb N(A,M)=\sum_j\sum_mN(A-2\sqrt{j(j+1)},M-m)
+\theta\big(A- 2\sqrt{|M|(|M|+1)}\big).\ee
All values of $M$ have to be admitted in the calculation
because the numbers $N$ with fixed $M$ do not close.
This relation is understood in a way similar to the earlier one.
The distributions of punctures are split into those with only
one puncture and those with at least two. If there is one
puncture, its $j$ is related to $A$, and whether it can have a
projection $M$ or not depends on whether $A\ge 2\sqrt{|M|(|M|+1)}$.
If there are two or more punctures, the spin $j$ of the first one
is considered and the remainder has $A$ reduced to 
$A- 2\sqrt{|M|(|M|+1)}$ and $M$ reduced to $M-m$.

The Fourier transform of the recursion relation involving $M$ is 
\bb \int{d\omega\over 4\pi}e^{-i\omega M}\tilde N(A,\omega) 
&=&\sum_j\sum_m
\int{d\omega\over 4\pi}e^{-i\omega(M-m)}
\tilde N(A-2\sqrt{j(j+1)},\omega)\nonumber\\  
&+&\theta\big(A- 2\sqrt{|M|(|M|+1)}\big).\ee
One sees that if $A$ is large, 
it is satisfied by
\bb \tilde N(A,\omega)\sim\exp(\lambda(\omega) A/2),\label{2}\ee
where $\lambda(\omega)$ obeys 
\bb 1=\sum_j\exp\Big[-\lambda(\omega)\sqrt{j(j+1)}\Big]\sum_me^{i\omega 
m}. \label{w}\ee 
For $\omega=0$, the equation for $\lambda(\omega)$ reduces to that for 
$\lambda$, so $\lambda(0)=\lambda$. This yields the dominant contribution $\exp
(\lambda A/2)$ seen above. For small $\omega$, $\lambda(\omega)$
falls quadratically, like $\lambda-c\omega^2$, say,
and the $\omega$ integral 
\bb N(A,0)\sim\int_{-2\pi}^{2\pi}{d\omega\over 4\pi}
\exp[\lambda(\omega)A/2]=\int_{-2\pi}^{2\pi}{d\omega\over 4\pi}
\exp[(\lambda-c\omega^2)A/2]\ee
becomes a gaussian, which is seen to be proportional to $A^{-1/2}$  for
large $A$ by scaling: \bb  N(A,0)\propto{\exp(\lambda A/2)\over
A^{1/2}}.\ee 
This indicates a correction $-\frac{1}{2}\log A$ \cite{bhe1} in the entropy,
which violates the area law.


Now we take into account the possibility of $M$ being
equal to zero only {\it modulo} $k/2$,
which is the actual requirement.
One may approximate the integral for spin projection $M$: 
\bb
N(A,M)= N(A)e^{-M^2/(2cA)}.
\ee
Since $M=0$ mod $k/2$, one has to sum over the values $M=rk/2$, where 
$r=0,\pm 1, \pm2,...,$ and there arises a factor
\bb
\sum_r e^{-r^2k^2/(8cA)}
\ee
which, on approximation by an integral over $r$,
is seen to involve a factor $\sqrt{A}/k$,
cancelling the square root in $N(A)$.
Thus,  
\bb
N_{corr}(A)= \frac{1}{k}\exp ({\lambda A\over 2}),
\ee
indicating that the log correction in $A$ is absent, but there is a correction
involving the {\it classical area} $k$.

More accurately, one has to add up
$ N(A)=\sum_r N(A,rk/2).$ This means
\bb N(A)=\int_{-2\pi}^{2\pi}\sum_r e^{-irk\omega/2}{d\omega\over 4\pi}\tilde N(A,\omega).\ee
The sum over $r$ of $e^{-irk\omega/2}$ is a geometric series.
For $k\omega=0 {\rm ~mod~}4\pi$,
it becomes a sum of terms all equal to unity and hence diverges.
For other values of $k\omega$, it vanishes because of oscillations.
This in fact produces $4\pi\sum_s\delta(k\omega-4s\pi)$,
where $s$ goes over all integers. 

Because of the bounded range of $\omega$,
$s$ takes a finite number of values:
\bb
N(A)\sim\frac{1}{k}\sum_{s=-k/2}^{k/2} \exp({\lambda(4s\pi/k)A\over 2}),
\label{3}\ee
implying that the entropy is a sum of a finite number of exponentials
and thus has no scope for a logarithmic correction in $A$ \cite{agpm}, 
though there is a correction $\log k$ involving the classical area $k$.
This is exponential in the area, but does
the correction violate the requirements?

We have to reformulate the requirement of an additive entropy.
If the number of states is allowed to depend on an area
and  an additional additive variable which we call $M$, the condition
(\ref{1}) gets modified to
\bb \int{d\omega\over 4\pi}e^{-i\omega M}\tilde N(A_1+A_2,\omega) 
&=& \sum_m\int{d^2\omega\over (4\pi)^2}e^{-i\omega_1 m-i\omega_2(M-m)}
\tilde N(A_1,\omega_1) \tilde N(A_2,\omega_2)\nonumber\\
&=& \int{d\omega\over 4\pi}e^{-i\omega M}
\tilde N(A_1,\omega) \tilde N(A_2,\omega).\ee
This is satisfied by (\ref{2}),
with no constraint on $\lambda(\omega)$.
With the help of
a condition of $M$ vanishing modulo $k/2$, we may reach (\ref{3}) again.
This yields a correction $\log k$ in the entropy.
Thus the $k$ correction obtained in loop quantum gravity
can be accommodated in the general
formalism through $M$.

\section{SU(2) loop quantum gravity}

In the alternative SU(2) formulation of loop quantum gravity \cite{k},
the number of states for a distribution of spins over punctures 
arises from properties of SU$_q$(2) as
\bb
N=\frac{2}{k+2} \sum_{a=1}^{k+1}\sin^2 {a\pi\over k+2}\prod_p 
\bigg[{\sin {a\pi(2j_p+1)\over k+2} \over\sin {a\pi\over k+2}}\bigg],
\ee
where the product is over punctures $p$.

Note that $N$ can be written as
\bb
N=\sum_{a=1}^{k+1} N_a,
\ee
where each $N_a$ depends on the angular momentum quantum numbers at the
different punctures:
\bb
N_a=\frac{2}{k+2}\sin^2\frac{a\pi}{k+2}\prod_p{\sin\frac{(2j_p+1)a\pi}
{k+2}\over\sin\frac{a\pi}{k+2}}.
\ee
This may be further split into
\bb
N_a&=&\frac{2}{k+2}\sin^2\frac{a\pi}{k+2}\bar N_a,\nonumber\\
\bar N_a&=&\prod_p{\sin\frac{(2j_p+1)a\pi}
{k+2}\over\sin\frac{a\pi}{k+2}}\equiv\prod_p[2j_p+1]_a.
\ee
With this notation, Meissner's relation looks like
\bb
\bar N_a(A)=\sum_j[2j+1]_a\bar N_a(A-2\sqrt{j(j+1)})+[\sqrt{A^2+1}]_a.
\ee
The last term comes from $2j_{max}+1$.
For large $A$, this leads to the solution
\bb
\bar N_a=\exp(\lambda_aA/2),
\ee
with 
\bb
\sum_{j=\frac12}^{k/2}[2j+1]_ae^{-\lambda_a\sqrt{j(j+1)}}=1.
\ee
Thus 
\bb
N=\sum_aN_a=\sum_a
\frac{2}{k+2}\sin^2\frac{a\pi}{k+2}\exp(\lambda_aA/2),
\ee 
is essentially a sum of terms each increasing exponentially
with $A$ and is dominated for large $A$ by the largest $\lambda_a$,
which is expected to occur for small $\frac{a}{k+2}$ as this makes
$[2j+1]_a$ largest for each $j$. Thus the dependence on $A$ is
purely exponential \cite{pm}. Note that
there is a $3{\log (k+2)}$ correction much like the U(1) case,
the number 3 taking into account the factor $\sin^2\frac{a\pi}{k+2}$,
which becomes $(\frac{a\pi}{k+2})^2$ for small $\frac{a}{k+2}$.
The coefficient 3 is different from the earlier calculations, 
but the ratio of U(1) and SU(2) remains the same.
This is related to the fact that the bigger group has three times as many
generators as the smaller group.


\section{Conclusion}

To summarize,
the laws of black hole mechanics suggested $S\propto A$.
The black hole radiation temperature indicated $S=\frac{A}{4\ell_P^2}$.
Loop quantum gravity has earlier indicated 
{$S=\frac{A}{4\ell_P^2}-\frac12\log A$}
and 
$S=\frac{A}{4\ell_P^2}-\frac32\log A$
in the U(1) and SU(2) formulations respectively.
Now we find 
{$S=\frac{A}{4\ell_P^2}-\log k$}
and {$S=\frac{A}{4\ell_P^2}-3\log (k+2)$}
in the U(1) and the SU(2) formulations respectively.
While the older corrections violate the area law, the newer results
are consistent with the multiplicativity of the number of states 
when allowed to depend on another variable besides the area.

\end{document}